\documentclass[conference]{IEEEtran}

\usepackage{textcomp}
\usepackage{amsmath}
\usepackage{stmaryrd}
\usepackage{textcomp}
\usepackage[short,nodayofweek,level,12hr]{datetime}

\begin{document}

\title{A Stricter Heap Separating Points-To Logic \\ \small{(Extended Abstract)}}

\author{
  \IEEEauthorblockN{Ren\'{e}~Haberland,
		    Kirill~Krinkin}
  \IEEEauthorblockA{Saint Petersburg Electrotechnical University "LETI"\\
  Saint Petersburg, Russia}
}

\maketitle

\IEEEpeerreviewmaketitle

\textbf{Keywords.}
separation logic, dynamic memory verification, points-to memory model, partial heap specification, ambiguous heap assertions.\\

\textbf{Abstract.}
Dynamic memory issues are hard to locate and may cost much of a development project's efforts as it was found e.g. in \cite{miller90} and was repeatedly reported similarly afterwards independently by different persons. Verification as one formal method may proof a given program's heap matches a specified dynamic behaviour.
Dynamic (or heap) memory, is the region within main memory that is manipulated by program statements like \textit{alloc}, \textit{free} and pointer manipulation during program execution. Usually, heap memory is allocated for problems where the amount of used memory is unknown prior to execution. Regions within the heap may be related ``\textit{somehow}'' with each other, often, but not always, by pointers containing absolute addresses of related heap cells \cite{Jones11}. The data structure described by all valid pointer variables manifests heap graphs.

A heap graph is a directed connected simple graph within the dynamic memory which may contain cycles, and where each vertex represents an unique memory address and every edge links two heap vertices. The heap graph must be pointed by at least one variable from the local stack or a chain of other heap graphs which is finally pointed by at least one stacked variable. Heap vertices may not overlap. A heap formula expresses the assertion on dynamic memory and can either be a heaplet, or a recursively defined heap-spatial or logical formula.

One of many different ways to specify the heap graph is by a points-to model as it was proposed by Burstall \cite{burstall72} and Reynolds \cite{reynolds02}. Essentially, both specify pairs \textit{loc $\mapsto$ val} of a location and its pointed by value, they differ in whether an address or an immediate value is used. The advantage of a points-to model over e.g. a shape region is \textit{locality}, which causes only local changes to the heap graph specification on changes due to its edge-centric view, and provides an intuitive style of how program statements correspond to heap graph. Burstall and Reynolds introduce a non-repetitive \textit{Separation Logic}, which weakens, for instance, constants which become in fact functions. The underpinning theoretically apparatus is a \textit{Substructural Logic} \cite{restall00}.

Initially, Reynolds proposed to use the ``,''-operator in order to describe heap graphs, which works fine for linked-lists or cactus-shaped heap graph, but which would imply a search for perfect matchings, if proceeded for heap graphs in general. Reynolds defines the set of heap graphs, but not a single heap graph which can only be defined approximately from his written objections as either a conjunction of existing heaps or a disjunction of (possibly connected) heaps.

Our Class-instantiated objects are considered as pointer generalisations, though an object pointer may point at the same time to more than one object. Motivated mainly by our believe in many cases arbitrary heap access by immediate addresses can and should be restricted by different modelling, it is first of all not primarily an expressibility issue for that particular case.
Motivated by Prolog predicates may be introduced to specify heap, which are naturally relational and are not classical functions. Since its semantic is relational, it makes them flexible, for instance when it comes to express general heap graphs. We call parameterized predicates \textit{abstract}. 

The main new idea behind this approach is to distinguish strictly in syntax and semantics between heap conjunction and disjunction. Therefore, algebraic rules are agreed which may eventually be used to define equalities, which then might be used to toggle a SMT-solver reducing simplifications more efficiently and consequently reduce bloated verification rules. It may be considered as a side-effect verification rulesets could be checked for completeness according to specified heap terms. The motivation behind making operators stricter is to undermine exceptional cases, which eventually will make calculations simpler.

Locations may be local variables, objects, and object fields. All locations in a heap graph must be unique. With the heap terms defined the next pointer-pointed quantities may be associated: 1:1, m:m, and m:1. However, 1:m is prohibited, except we understand as an object with all outgoing pointers as ``\textit{one}''.  W.l.o.g. inner objects must always be modelled as exterior objects. Heap conjunction binds stronger than disjunction.
 Late binding is currently ignored. Pointers of pointers are also not further considered, although not prohibited, for the reason that they do not change anything essentially to a heap graph one could not do without further indirection. $\underline{true}$ interprets true for any matching heap ($\underline{false}$ in analogy), $\underline{emp}$ interprets only true if the matching heap is empty -- those are used for partial heap specifications.

The proposed heap conjunction says that two heaps are connectible, where the right heap must be a points-to expression, iff there exists exactly one joining point otherwise it interprets as false. False, after all is not undefined, and therefore conjunction on unconnectible heaps is total. Alternatively, the conjunction may be refined further which part is going to be source and which is going to be target. The above heap conjunction can be generalised according to the extended heap terms. By convention it is agreed that for $H1$ being a heap $H1 \circ \underline{emp}$ = $\underline{emp} \circ H1$ = $H1$ holds. When dealing with object fields, we agree further object accessors are left-associative $object1.field1.field2.field3$ = $((object1.field1).field2).field3$, so left parts of paths may be assigned by symbols: this is most important when using abstract predicates, because only fields for an unspecified object are usually provided.

All ``$\circ$''-connected heaps generate a commutative group with several arrangements: closure follows from totality of ``$\circ$'', identity is $\underline{emp}$, associativity holds only for connectible elements -- if heaps are not connectible, then \textit{false} is returned except it will be connected until last element of ``$\circ$'' is consumed. It always holds that unrelated heaps may not be ``$\circ$''-conjuncted, so $a \mapsto b \ \circ \ a \mapsto d = false$ regardless if $b=d$ holds or not. It does not matter in which order the (connectible) heaps are connected, important is that they are all connected, this establishes confluence of ``$\circ$''-joined heaps. Without any extra costs in Prolog, locations may be symbols. Due to associativity and closure the problem of abstract predicates may be interpreted as Word-problem.
Furthermore, existence of an inverse heap w.r.t. ``$\circ$'' always exists due to the generalised heap inversion $G \circ G^{-1}=\underline{emp}$, which can be shown over the heap term inductively. The possibility to refer to an inverse may be useful when a proof refutes in order to digest expected from actual heaps, and of course, it is as for instance Galois field extensions, a convenient technique in order to calculate in terms of algebraic equations. Intuitively inversion may be interpreted as heap negation plus some extra clean-up heap vertices, if those are no more needed. Let the convention be $\underline{emp}^{-1} = \underline{emp}$. It was found, that required precautions on calculations with inverses can easily be linearly adapted after each calculation step over heap conjunction and disjunction. First, this is the case whenever source/target are still in use, then they may not be substituted by disjunction. Second, when a bridging edge between graphs is removed, then a heap conjunction needs to be turned into a disjunction.  $(G_1 \circ G_2)^{-1} \equiv G_1^{-1} \circ G_2^{-1}$ holds for any heaps $G_1$ and $G_2$.

Heap disjunction becomes very straight and intuitive, because there are no more exclusive cases to be taken into consideration when defining verification rules: Two heaps $H1$ and $H2$ are indeed independent, iff $H1 \| H2$. Similar to heap conjunction, but under different circumstances, heap disjunction and heap partitions form a group. In analogy to point-wise heap conjunction, the disjunction of heaps may be point-wise, too -- both operations are dual.

Now heaps may be built upon consistent and strict operations. Besides these operations, heaps may also be used to define partially ordered sets with an infimum element $\underline{emp}$, a totally connected graph as supremum, and ``$\circ$''- operator as joining operator.

Local variables as points-to expressions, may be grouped together among ``$\circ$''-conjuncted heaps to invariant parts when procedures or methods are called, however it needs to be taken into consideration heap inversion may invalidate the \textit{heap frame rule} which may need make procedure calls specially aware on unaffected changes. Object fields are ``$\circ$''-conjuncted too, and so the same constant formulea are applicable to objects – just to distinguish all locals from all fields from a particular object, constant formulea, like $\underline{true}$, are parameterized for objects.

So, $a.f1 \mapsto x \circ \underline{true}(a)$ may denote property $f1$ of object $a$ on the lefthand-side of ``$\circ$'' where $\underline{true}(a)$ may denote all remaining fields from $a$ (except $f1$). This is why, in $\underline{true}(a) \circ \underline{true}(a)$ the lefthand side accumulates all properties where the second $\underline{true}(a)$ actually accumulates none which makes it equals to $\underline{emp}$. The bottom line is, the object fields need to be traced while verification, so partial specification may fill in all non-specified fields automatically, which eases specification a lot without making the specification imprecise by default. From the point of view of \textit{Separation of Concern} it is highly recommended abstract predicates specify as much as possible of an object's behaviour rather than spreading object behaviour all over different abstract predicate definitions -- also but not only because the stack-based approach, would be limited to abstract predicates calls in depth rather than in width.

The Object Constraint Language (OCL) \cite{oclspec} is a specification language for class-instantiated objects in companion to the Unified Modeling Language. It implements a considerable part of first-order predicate logic, furthermore it has got quantification, supports collection types and ad-hoc polymorphism by subclassing. There is a  way to specify an object's life-cycle and class methods. However, OCL does not know of pointers nor aliases. In combination with abstract predicates the new logic presented may be used as recommendation for an update of the recent OCL definition w.r.t. the intrinsic points-to model, which would benefit in better modularity and improved Separation of Concerns.

\section*{Acknowledgement}

Parts of this paper are prepared as a contribution to the
state project of the Board of the Ministry of Education of the Russian Federation (task \# 2.136.2014/K).


\begin{thebibliography}{1}
\bibitem{Jones11}
R.~Jones, A.~Hosking, and E.~Moss, {\em The Garbage Collection Handbook: The
  Art of Automatic Memory Management}.
\newblock Chapman \& Hall/CRC, 2nd~ed., 2011.

\bibitem{burstall72}
R.~M. Burstall, ``Some techniques for proving correctness of programs which
  alter data structures,'' in {\em Machine Intelligence} (B.~Meltzer and
  D.~Michie, eds.), vol.~7, pp.~23--50, Scotland: Edinburgh University Press,
  1972.

\bibitem{reynolds02}
J.~C. Reynolds, ``Separation logic: A logic for shared mutable data
  structures,'' in {\em Proceedings of the 17th Annual IEEE Symposium on Logic
  in Computer Science}, (Washington, DC, USA), pp.~55--74, IEEE Computer
  Society, 2002.
  
\bibitem{restall00}
G.~Restall, {\em Introduction to Substructural Logic}.
\newblock Routledge, 2000.
\newblock ISBN 041521534X.

\bibitem{miller90}
B.~P. Miller, L.~Fredriksen, and B.~So, ``An Empirical Study of the Reliability of UNIX Utilities,'' in {\em Proc. of the Workshop of Parallel and Distributed Debugging}, pp.~1-22, Digital Equipment Corp., 1990.

\bibitem{oclspec}
\em{Object Management Group (OMG)}: Object constraint language version 2.2,
  available at http://www.omg.org/spec/ocl/2.2, from 15th Feb 2010.
\end{thebibliography}
\end{document}